\documentclass[times]{smrauth}
\pdfoutput=1
\usepackage{etex}

\usepackage[utf8]{inputenc}
\usepackage[american]{babel}
\usepackage[boxed,ruled,vlined]{algorithm2e}

\usepackage{todonotes}

\usepackage{paralist}
\usepackage{graphicx}
\usepackage{url}
\usepackage{multicol}
\usepackage{xspace}
\usepackage{hyperref} 
\usepackage{subfig}
\usepackage{booktabs}
\usepackage{amsmath,amsfonts,amssymb}

\usepackage{float}
\floatstyle{plaintop}
\restylefloat{table}

\SetKwProg{Fn}{Function}{ is}{end}

\newcommand{\vppo}[1]{}

\newcommand{\algoshortname}{GD-GNC\xspace}

\newcommand{\graphid}{G_{\nodesetid,\edgesetid}}
\newcommand{\nodesetid}{\mathcal{N}}
\newcommand{\edgesetid}{\mathcal{E}}
\newcommand{\nbiter}{N}
\newcommand{\nbsubjects}{50\xspace}
\newcommand{\probgnc}{p}
\newcommand{\probdouble}{q}
\newcommand{\nthnode}[1]{n_{#1}}
\newcommand{\currentnode}{\nthnode{i}}
\newcommand{\randomnode}{\nthnode{j}}
\newcommand{\takerandomnode}{Select uniformly at random a node $\randomnode$ in $\graphid$}

\newcommand{\edgeitem}[2]{(#1  #2)}

\newcommand{\maxksoperator}{\delta}

\newcommand{\NullAltHyp}[2]{\hspace{5mm}$H_0$: #1.

\hspace{5mm}$H_1$: #2.}

\newcommand{\etal}{\textit{et al.}}
\newcommand{\eg}{\textit{e.g.}\xspace}
\newcommand{\ie}{\textit{i.e.}\xspace}
\newcommand{\etc}{\textit{etc.}\xspace}
\newcommand{\cf}{\textit{cf.}\xspace}
\newcommand{\aka}{\textit{a.k.a.}\xspace}

\newcommand\RQOneCommonShape{RQ1. Are there common structures of node degree distributions in software dependency graphs?\xspace}

\newcommand\RQTwoGNCvsBaxter{RQ2. How does \algoshortname\ compare to other generative models? We consider Baxter \& Frean's model which is the best model up to date.}

\begin{document}
	\runningheads{V. Musco et al.}{GD-GNC}

	\title{A Generative Model of Software Dependency Graphs to Better Understand Software Evolution}
	
	\author{Vincenzo Musco,
		Martin Monperrus and Philippe Preux\\
		University of Lille and Inria, France}
	 
	\begin{abstract}
      Software systems are composed of many interacting elements. A natural way to abstract over software systems is to model them as graphs. 
      Being an evolving system, a program, hence its dependency graph, evolves along time, mostly by growing, that is increasing its number of nodes, and its number of edges. In science, it is a recognized approach to model how an object evolves. This let us hypothesize and study the rules of evolution at work while a program is developed.
      We propose GD-GNC, a new generative model of synthetic software dependency graphs. This model is used to generate synthetic graphs aiming at being similar to graphs of real software.
      We study a set of important properties of graphs: degree distribution, spectra, diameter, transitivity, modularity, average shortest path length. We compare different generative models and show that GD-GNC produces synthetic graphs that have lots of similarities with the graphs of real software. 
From a software development perspective, this model hints that remix of existing classes to create new features is an important rule of software evolution.
      From both an empirical and an experimental point of view, having a good generative model of software dependency graphs opens the way to many studies of software and its evolution.
    \end{abstract}

	
  \maketitle

\section{Introduction}
Software systems are composed of many elements interacting with each other. For instance, there are hundreds of thousands of interconnected functions in a Linux kernel \cite{BowmanLinux1999}.
A natural way to abstract over software systems is to model them as graphs \cite{BaxterSoftware2008,BhattacharyaGraph2012,ConcasSuitability2006,LouridasPower2008,ValverdeLogarithmic2005}. 

In this paper, we consider software dependency graphs of object-oriented software, where each node represents a class and each edge corresponds to a compile-time dependency. We study the structure of those graphs through their degree distributions, that is the distribution of the number of edges coming from and to a node. Based on the analysis of 50 software systems written in Java, we show that there exists different software systems that have similar degree distribution. 
This is a surprising result: despite being developed by different persons with different processes in different domains, a common pattern emerges concerning their degree distribution.
This makes us hypothesize that whatever the development design, the structure of software dependency graphs tend to be similar.

In network science, a generative model defines a set of rules used to synthesize artificial graphs\footnote{In this paper, we use ``network'' and ``graph'' interchangeably.} in a given domain. For example, there exists generative models that aim at producing graphs that are \textit{e.g.} similar to the World Wide Web \cite{KleinbergWeb1999}. In software engineering, such a model would generate graphs which look like real software graphs. 

In this paper, our goal is to propose a generative model of software dependency graphs.
If this model produces graphs that fit the empirical data, it would mean that the generative operations are good candidates to describe the core operations which result in software graphs' structure.
In other words, a good generative model may encode the evolution rules that are behind the structure of software systems. 

Our experimental methodology answers two research questions.
We devise an experimental methodology to answer the following research question:
\RQOneCommonShape To address this question, we compute the degree distributions of 50 Java applications and look for commonalities.

We propose a generative model of software graphs called GD-GNC and we address the second research question: \RQTwoGNCvsBaxter~
First, we evaluate its capacity to create graphs whose degree distribution is close to the empirical ones observed in real software systems and compare its fit-to-data to the closest model of the literature \cite{BaxterSoftware2008}. Then, we also compare the graphs generated with GD-GNC and the real graphs through different graph metrics: metrics related to the length of paths to reach a node from an other one, metrics related to the community structure of graphs, degree distributions compared in different ways, and spectral properties.

To sum up, our contributions are:

\begin{itemize}
  \item empirical evidence of a common asymmetric structure of dependency graphs in object-oriented software systems written in Java: the in-degree distribution is different from the out-degree distribution;
  \item a generative model of software dependency graphs;
  \item the validation of the model regarding its ability to fit 50 graphs of real software systems totaling 23,178 nodes and 108,404 edges;
  \item a speculative explanation of the evolution rules of software.
\end{itemize}

The source code is publicly available on Github \footnote{\url{https://github.com/v-m/GDGNC}}.

The rest of this paper is structured as follows. Section~\ref{section:definitions} defines the main concepts used in this paper. In Section~\ref{section:shape}, we look closer at real software dependency graphs and highlight their common structure. In Section~\ref{section:GD-GNCmodel}, we introduce a new generative model  for software dependency graphs and we analyze its fitness.  In Section~\ref{section:discussion}, we discuss our findings from a software engineering perspective. In Section~\ref{section:relatedworks}, we present works related to this contribution. In Section~\ref{section:conclusion}, we conclude this paper.

\section{Background}

In this section, we provide background knowledge about the concepts used in this paper.

\subsection{Graphs, degrees, distributions, and spectra}
\label{section:definitions}

Let $G(N,E)$ be a directed graph composed of $N$ nodes and $E$ directed edges. Nodes are numbered from 1 to $N$ and each node is identified by this number. An edge $e$ connecting node $n_i$ to node $n_j$ is expressed as an ordered pair: $e = (n_i, n_j)$, where $e \in E, n_i, n_j \in \{ 1, ..., N \}$.

The \emph{in-degree} and the \emph{out-degree} of a node $n_i$ are respectively the number of edges going to $n_i$ (\ie{} the number of edges $(., n_i)$), and the number of edges leaving $n_i$ (\ie{} the number of edges $(n_i, .)$). We use the term \emph{degree} to refer to both of those concepts, and the terms \emph{in-degree} and \emph{out-degree} when the distinction is necessary.

The \emph{degree distribution} of a graph is the proportion of nodes of each degree in that graph. It sums to 1. In this paper, we always consider \emph{cumulative distribution} functions (CDF) of degrees, \ie{} the proportion of nodes whose degree is smaller or equal to a given value. Let $deg (n_i)$ denote the degree of node $i$, for any non negative integer $d$, $CDF (d)$ is defined as follows:

$CDF (d) = \frac{| \{ n \in \{ 1, .., N \}, deg (n) \le d \}| }{N}$ where $| S |$ denotes the size of the set $S$.

Let $deg_{max}$ denote the maximal degree of any node in the graph; $\sum_{d\le{}deg_{max}} CDF (d) = 1$. 
$CDF_{in}$ and $CDF_{out}$ are defined according to the in-degree, and out-degree of nodes.

Noncumulative distributions are to be avoided as they are sources of mistakes \cite{LiToward2005}. Cumulative distributions are more appropriate to analyze noisy and right-skewed distributions \cite{NewmanStructure2003}, which is the type of distributions we face. 
The related literature also considers \emph{inverse cumulative distributions} (ICD). We will do the same for the sake of easy comparison with prior works. For any non negative integer $d$, $ICD$ is defined as follows:

$ICD (d) = 1 - CDF (d)$

A graph made of $N$ nodes may be represented by its adjacency matrix $A$ of size $N\times{}N$ where the entry $A_{i,j}$ is equal to 1 if there exists an edge from node $n_i$ to $n_j$, 0 otherwise. The adjacency matrix of a non directed graph is symmetric, whereas that of a directed graph is not symmetric. The set of eigenvalues and eigenvectors of the adjacency matrix provides important characteristics of the related graph: eigenvalues and eigenvectors come in pairs (so-called ``eigenpairs''): an eigenvector $v$ is associated to each eigenvalue $\lambda$ and an eigenpair satisfies $A v = \lambda v$. There are $N$ eigenpairs; the set of eigenvalues is known as the spectrum of the matrix; each of the $N$ eigenvectors has $N$ components. The eigenvector associated to the largest eigenvalue is known as the ``network value''.
When the graph is not directed, the adjacency matrix is symmetric and its eigenvalues are real, and all components of its eigenvectors are real. When the graph is directed, the adjacency matrix is not symmetric and eigenvalues are in general complex numbers, and eigenvectors have complex components \cite{BRUALDI20102181}.

There is a host of properties that describe various aspects of a graph, the relation between these properties being more or less (often less) understood. Let us mention the properties we consider in this paper:

\begin{itemize}
\item the diameter of the graph which is the length of the longest shortest path between any pair of nodes.
\item The average shortest path length which is the average length of the shortest path between any pair of nodes.
\item The transitivity (or clustering coefficient) and the modularity are two different measures of whether there exists some subsets of nodes which are more connected to each others than the average. Though the exact relation between these two metrics is not yet clear, there are some similarities \cite{DBLP:journals/corr/abs-1207-3234}.
\item About the degrees of nodes: the distribution of these degrees, and the coefficient of the law that describes it: power law and log-normal distributions are considered. For directed graphs, we distinguish the in-degree distribution, and the out-degree distribution.
\item Spectral properties: we consider the spectrum as well as the network value.
\end{itemize}

Node degrees and their distributions are basic properties which are directly influenced by many properties of a graph. 
As others\cite{BaxterSoftware2008,BhattacharyaGraph2012,LouridasPower2008,ValverdeLogarithmic2005, PotaninScalefree2005,MyersSoftware2003,KrapivskyNetwork2005}, one important assumption of our work is that the cumulative degree distribution of nodes reflects an essential aspect of the software structure.
However, we do not claim that a good fit in degree distribution for a given generative model necessarily means that the model's assumptions hold in practice.

Despite its simplicity, its degree distribution is strongly characterizing a graph. It is true that the properties mentioned above are neither clearly, nor uniquely related to this distribution. It is also true that given a degree distribution, the number of graphs having this degree distribution may be very large. However, the degree distribution is very specific to a graph, in particular because the graphs have log-normal degree distributions. For the sake of clarity and to keep the paper within reasonable length, we will mainly focus on degree distributions; however, we have studied and we will report the key observations that we have made regarding all the properties mentioned above.

\subsection{Software Dependency Graphs}
\label{text:depgraphs}
A large variety of graphs can be derived from software, each one focusing on particular characteristics. Hence, nodes and edges can have various meanings. An example of software graph is the \textit{dependency graph} in which nodes are modules (\eg{} packages, classes, \etc{} depending on the chosen granularity) and edges reflect that an element uses another one (\eg{} function call, inheritance, field access, \etc{}).

\label{text:softgraphsaredirected}
Dependency graphs are directed graphs as dependencies are oriented. The nodes composing a dependency graph can be of two different types. First, there are \emph{application nodes} (\aka app nodes) that are nodes which belong to the core software itself. Second, there are \emph{library nodes} (\aka lib nodes) which are nodes which belong to an external library. 
For instance, a class of software package ``Eclipse'' may use a class in\texttt{java.util} library. The former is an application node, the latter is a library node.

\begin{figure}
	\centering
	\includegraphics[width=12cm]{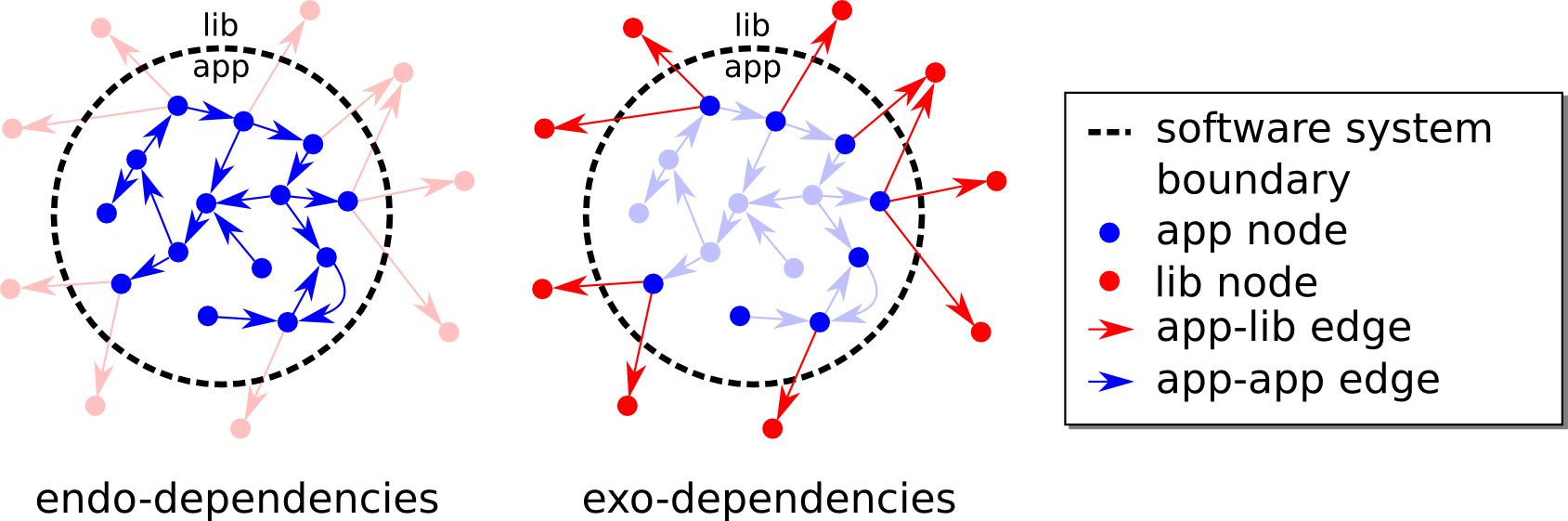}
	\caption{Illustration of the set of endo-dependencies (left) and exo-dependencies (right) in a dependency graph.}
	\label{figure:applibschema}
\end{figure}

From this distinction between the two types of nodes, there are two types of edges in a software dependency graph: 
\begin{inparaenum}[(i)]
\item [an app-app edge] connects an application node to an application node. We call them \emph{endo-dependencies}: they express that a core class depends on another core class;
\item [an app-lib edges] connects an application node to a library node. We call them \emph{exo-dependencies}: they express that a core class depends on a library class.
\end{inparaenum}
Fig.~\ref{figure:applibschema} illustrates these notions. On the left-hand side of the figure, endo-dependencies are emphasized; on the right-hand side, the red arrows crossing the system boundary are exo-dependencies.
\emph{In this paper, we exclusively focus on endo-dependencies.}

\subsection{Generative Models}

A generative model for graphs is an algorithm that generates graphs. A generative model takes a set of parameters as input (such as the number of nodes or a threshold).
A generative model may be \emph{deterministic} or \emph{stochastic}. For a given set of parameters, a deterministic model always generates the exact same graph whereas a stochastic model generates a new graph each time it is run.

There are a few major categories of generative models and we introduce only the two most important, at least from our perspective. In particular, we do not discuss models that generate graphs with a given degree distribution; these models are not growing graphs, and hence can not explain rules that govern graph growth. One may check \cite{ChakrabartiGraph2006} for further models. The simplest and earliest is the model independently introduced by Gilbert, and Erd\"{o}s and R\'enyi in 1959, both known as Erd\"{o}s-R\'enyi models: ER$(n,m)$ generates a random graph with $n$ nodes and $m$ edges chosen uniformly at random, and ER$(n,p)$ generates a random graph with $n$ nodes and any two nodes being connected with a probability $p$. These are very nice theoretical models, but no graph observed in the real is ER: they have properties that real graphs do not have, and it is not possible to generate ER graphs that share the properties of real graphs. The other important category we mention are the graphs generated following a preferential attachment principle: the graph is generated incrementally, adding one node at each iteration, and connecting this node to already existing nodes following the idea that the more connections a node already has, the more likely it is to be connected to the new node (``rich gets richer'' principle). Though a relatively old idea dating back at least to 1925, this sort of graphs have gained considerable attention these last 20 years when it was observed that many real graphs may be described by such a preferential attachment process. The most well-known model of this sort is the Barabasi-Albert model \cite{ba}. Among other properties, such graphs have the degree distribution of the nodes that usually follow a power law or a log-normal law; there is some controversy on this point \cite{newman,lognormal,Downey2005790}, also in the field of software engineering \cite{MullenGokhale2005,MullenGokhale2006}. In contrast, the degree distribution is binomial for ER graphs. In the case of a power law, the proportion of nodes of degree $k$ scales like $k^{-\gamma}$, the parameter of the distribution $\gamma$ being typically around 3. These graphs are also known as scale-free. In the case of a log-normal law, the logarithm of this proportion scales like $e^{-\frac{(\ln{k}-\mu)^2}{2\sigma^2}}$. The diameter and the average shortest path length scale as $\frac{\ln{n}}{\ln{\ln{n}}}$ where $n$ is the number of nodes of the graph \cite{Chen20081405}. As log-normal distribution is continuous, a discrete log-normal law has been proposed under the name discrete Gaussian exponential (DGX) \cite{Bi:2001:DDM:502512.502521}.

In our case, we generate graphs that are intended to be similar to software dependency graphs.
We consider two types of graphs: those resulting from an analysis of software systems, and those created by a generative model. The former are qualified as ``empirical'' or ``true'', the latter being qualified as ``synthetic'' or ``artificial''.

\section{Study of the Common Structure of Software Dependency Graphs}
\label{section:shape}

We want to determine whether there exists structures shared by different software applications. As the production of those software packages is influenced by different factors (management and development teams, development techniques, etc.), it is \textit{a priori} expected that there is no common structure. To the opposite, finding common structures would be an interesting fact as it would mean that there exists common evolution mechanisms shared across application domains and development styles.
Hence, our first research question is:

\textbf{\RQOneCommonShape}

\subsection{Protocol}

Our protocol consists of a technique to extract graphs from software code, applied to a given dataset, and a statistically sound analysis method.

\subsubsection{Dependency Graph Extraction}

We consider object-oriented software written in Java. We focus on the \emph{class granularity} (\ie{} one node represents one class), as this is the most important modularity unit in object-oriented software.

This extraction is performed by Dependency Finder\footnote{\url{http://depfind.sourceforge.net/}}.
This mature and open-source tool takes as input Java byte code and outputs all dependencies being found. In this section, graph metrics are computed using the NetworkX\footnote{\url{http://networkx.lanl.gov/}} library.
The dependencies are computed on the main source; test cases are not considered. We use DependencyFinder in class mode (hence discarding package-level dependencies). 

As already mentioned, we only consider endo-dependencies, that is, edges connecting internal nodes of the project to each others and not those connecting to external libraries (\cf{} section~\ref{text:depgraphs}), because we aim at understanding the inner structure of software graphs, regardless of the number of libraries that are included and the amount of calls to library functions.

The discrimination between endo-dependencies is done using a filter on the package of the source and destination of the dependency edge. For each subject, we define a package prefix (e.g. ``org.myapp''), and if both the source and destination fully-qualified classes start with this prefix, the edge is considered endo-dependent, or it is considered exo-dependent.

\subsubsection{Dataset}
\label{section:dataset}

To determine an acceptable dataset size, we examined the dataset size in related publications: they range from 1 to 12 \cite{BaxterSoftware2008,BhattacharyaGraph2012,LouridasPower2008,ValverdeLogarithmic2005, PotaninScalefree2005,MyersSoftware2003,KrapivskyNetwork2005}.
We aim at building a dataset that is at least at big as those used in previous work. Also, we aim at reusing a peer-reviewed dataset of Java applications in order to mitigate the risk of cherry-picking.

SF100 \cite{SF100} is a dataset that meets our requirements\footnote{The dataset can be downloaded at \url{http://www.evosuite.org/files/SF100-20120316.tar.gz}}.
It contains 100 Java applications, given as Jar files containing the classes in Java bytecode.
In a pilot experiment, we realized that SF100 contains many Java projects that are too small (a few classes only). To observe the structural properties we are interested in, a certain size of graphs is necessary. Consequently, in our experiments, we consider the 50 largest projects of SF100, that is all projects containing at least 56 classes.

\subsubsection{Comparison of Degree Distributions}
\label{section:ksstat}
We want to compare the degree distributions of a set of graphs, \ie{} software.
To that end, we use 
the Kolmogorov-Smirnov statistic $K$ given by eq.~\eqref{formula:metricks} in which $sup$ is the supremum of a set, $F_1$ and $F_2$ are the cumulative of two distributions to compare and $x$ ranges over degree values.

\begin{equation}
  K_{F_1, F_2} = \sup\limits_{x}|F_1(x) - F_2(x)|
  \label{formula:metricks}
\end{equation}

$K$ is a numerical value that indicates how close the two distributions are: the lower $K$, the closer the distributions. $K$ does not depend on any hypothesis made on the distribution: the Kolmogorov-Smirnov test is non parametric.

\subsection{Results}

\begin{figure}
  \includegraphics[width=75mm]{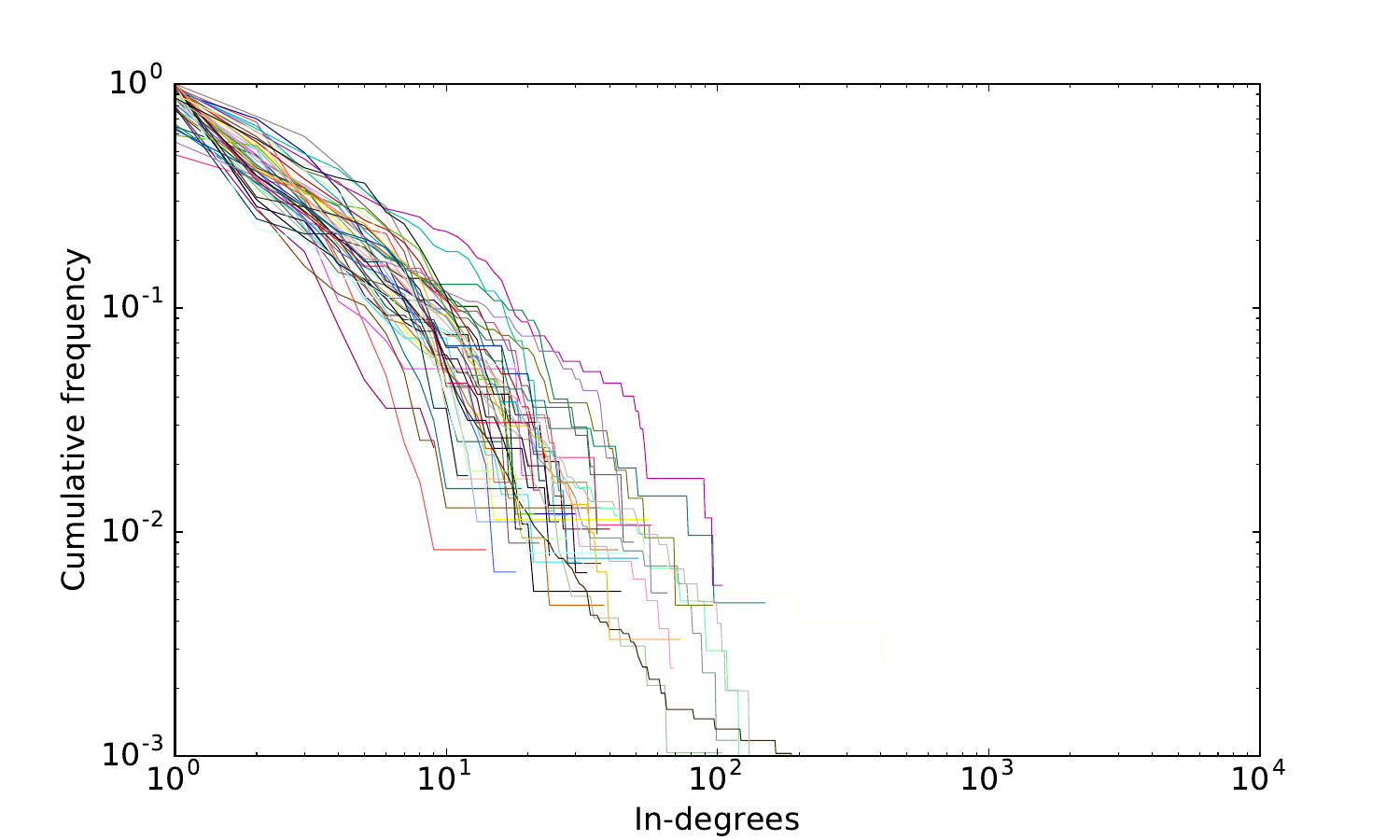}
  \includegraphics[width=75mm,page=2]{figures/softwareshapeplot50new.pdf}
  \caption{Inverse cumulative in- and out-degree distribution for the \nbsubjects software applications of our dataset (axes on a logarithmic scale). }
  \label{figure:softshape}
\end{figure}

Figure~\ref{figure:softshape} shows the plot of the in-degree (left) and out-degree inverse cumulative distributions (right) for our dataset. The scale is bi-logarithmic.
There is a different line and a different color for each software of the dataset.
We make two observations.

First, the in-degree distribution looks different from the out-degree distributions. The out-degree ICD is more bended, while the in-degree ICD is straighter.
This observation has already been made in previous works \cite{ValverdeLogarithmic2005,MyersSoftware2003,ChalletBug2004}: in-degree distributions and out-degree distributions are different.
As mentioned above, some have seen power laws in these distributions; we would rather follow those who see log-normal distributions, or its discretized version, the Discrete Gaussian Expoential (DGX). This controversy has already been studied \cite{LouridasPower2008,PotaninScalefree2005} and thus remains out of the scope of this paper.

Second, the position and the shape of distributions are graphically similar, this indicates that there are common structures across software applications. 
In order to assess this observation in a statistical manner, we now express our null hypothesis and the alternate hypothesis:

\NullAltHyp{Samples from the software in-degree distributions (resp.\@ out-degreee distributions) are drawn from the same distribution}{Samples from the software in-degree distributions (resp.\@ out-degree distributions) are not drawn from the same distribution}

Using the \emph{two-sample Kolmogorov-Smirnov test} on each pair of degree distributions of the dataset, we can statistically assess the presence of a  similar structure across software applications in our dataset. Based on the $K$ statistic, one decides on rejecting or not $H_0$ according to a given confidence level. If $H_0$ is not rejected, we gain confidence about the common structure for those two software. On the other hand, if $H_0$ is rejected, the test outcome can not be used to conclude about the common structure (which does not necessarily mean that the two software graphs are not drawn from the same distribution).

Table \ref{table:h1h0reject} gives the results of running 2,450 two-sample Kolmogorov-Smirnov tests with a confidence level $\alpha$ of 0.01
(we need to test each pair of software, hence $C^2_{\nbsubjects} = 1225$ tests, which is doubled ($2 \times 1225 = 2450$) since we test in-degrees and out-degrees). 
In this table, the rows provide the results for in-degree, out-degree, and both distributions. The second and third columns provide the number and the ratio of tests for which the two-sample Kolmogorov-Smirnov test has rejected $H_0$. The fourth and fifth columns provide the information when $H_0$ can not be rejected.

As we can see, for 79\% of the tested pairs, the common distribution hypothesis cannot be rejected. However, this affirmation does not necessarily involve that there is a unique distribution shared by all those software. On the other hand, for the remaining 21\% of tested pairs for which $H_0$ is rejected at this confidence level, no conclusion can be drawn.

\begin{table}
  \centering
  \begin{tabular}{lrrcrr}
    \toprule
    & \multicolumn{2}{c}{\textbf{$H_0$ Rejected}} & \phantom{abc} & \multicolumn{2}{c}{\textbf{$H_0$ Not rejected}} \\
    \cmidrule{2-3} \cmidrule{5-6}
    & Count & Ratio & \phantom{abc}& Count & Ratio \\
    \midrule
    In              &         208/1,225  &         17\%  &&         1,017/1,225  &         83\% \\
    Out             &         300/1,225  &         24\%  &&           925/1,225  &         76\% \\
    \midrule
    \textbf{Both}   & \textbf{508/2,450} & \textbf{21\%} && \textbf{1,942/2,450} & \textbf{79\%} \\
    \bottomrule
  \end{tabular}
  \caption{Number of times the $H_0$ hypothesis is rejected or accepted for in-degree, out-degree and both CDFs according to the two-sample Kolmogorov-Smirnov test with a confidence level $\alpha$ of 0.01.}
  \label{table:h1h0reject}
\end{table}

\subsection{Summary}

For the majority of subjects, there is not enough evidence to reject the hypothesis of a common structure.
What is the reason behind this common structure? 
It is not due to the fact that all applications were developed by the same team. They were indeed developed by different people with different background from all over the world. In this paper, we explore a specific assumption: the way people evolve software is similar across projects, and as a result, software applications eventually share a common structure.

\section{A Generative Model for Software Dependency Graphs}
\label{section:GD-GNCmodel}

In this section, we present a new generative model of software dependency graphs. This model generates an arbitrary number of artificial dependency graphs. It is parametrized by three values: the expected number of nodes and two probabilities.

\subsection{Assumptions}
\label{sec:assumption}
Our model is built on three assumptions on how software evolution works.

\textbf{Assumption \#1 (increment):} When creating new features (new classes), they are built on top of existing classes. Hence when a new node is added to the generated graph, it is directly connected to existing nodes.

\textbf{Assumption \#2 (remix):} New features (new classes) depend on classes that are designed to be used together or at least that are compatible with each others.

\textbf{Assumption \#3 (refactoring):} Developers sometimes identify reusable units. In that case, they create a new class. This new unit of reusable functionality is used later on.

\subsection{The Generalized Double GNC Algorithm (\algoshortname)}

\begin{figure}
  \centering
  \includegraphics[width=10cm]{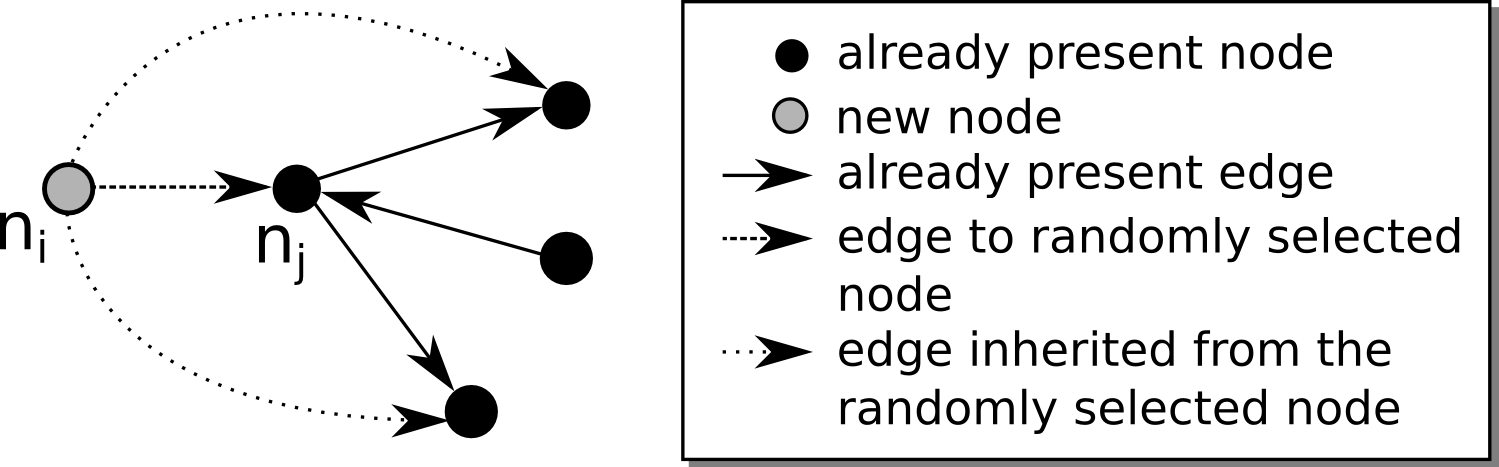}
  \caption{Illustration of GNC-Attach, the GNC primitive operation. The grey node $\currentnode$ is a new node added to the graph using the GNC primitive. The central node $\randomnode$ is selected uniformly at random and a directed edge is added from the new node to it (dashed edge). Then, a directed edge is also added from the added node to all successors of $\randomnode$ (dotted edges).}
  \label{figure:gncillustration}	
\end{figure}

\SetKwFunction{FGNC}{GNC\_Attach}

\begin{algorithm}
  \KwIn{$\graphid$ the digraph to which a new node $\currentnode$ is going to be added. $\graphid$ is composed of two elements: the set of existing nodes ($\nodesetid$), and the set of existing directed edges ($\edgesetid$).}
  \KwOut{$\graphid$ has been updated with a new node, and a set of new directed edges.}
  \Fn{\FGNC{$\graphid, \currentnode$}}{
    \takerandomnode\\
    \If{there is not yet an edge from $\currentnode$ to $\randomnode$}{
    Add an edge from $\currentnode$ to $\randomnode$ \\
    \For{all edges $\edgeitem{\nthnode{j}, \nthnode{d}}$ leaving $\randomnode$}{
      Add an edge from $\currentnode$ to $\nthnode{d}$\\
    }
  }}
  \caption{GNC-Attach Algorithm.\label{algo:gncclassic}}
\end{algorithm}

We name our generative model of software dependency graphs \algoshortname. It generalizes the GNC model \cite{KrapivskyNetwork2005}: this model is an iterative algorithm where, at each iteration, a new node $\currentnode$ is added to the graph and connected at random to a set of already existing nodes. In GNC, an existing node $\randomnode$ is selected according to a uniform distribution and directed edges are created from the new node $\currentnode$ to this node $\randomnode$ along with all its successors. This ``GNC-Attach'' primitive is illustrated in Figure~\ref{figure:gncillustration}.
Algorithm~\ref{algo:gncclassic} shows the core primitive for attaching nodes using GNC. 
GNC requires one parameter: the number of nodes of the resulting graph. 
It executes $n$ times the core function to create a graph with $n$ nodes.

\textbf{GD-GNC} implements the three assumptions presented in Section \ref{sec:assumption} in an algorithmic way.
Its pseudo-code is shown in Algorithm~\ref{algo:dblegncalgo}.
It consists in a main loop which at each iteration:
\begin{itemize}
  \item adds a new node $\currentnode$ to the existing graph (\textbf{Assumption \#1 -- increment}),
  \item selects an existing node $\randomnode$ uniformly at random,
  \item adds edges leaving $\currentnode$.:
  \begin{itemize}
    \item with probability $\probgnc$, $\currentnode$ is connected to $\randomnode$ in the same way as in the GNC-Attach algorithm (\ie{} a directed edge is created from $\currentnode$ to $\randomnode$ and from $\currentnode$ to each successor of node $\randomnode$).
With probability $\probdouble$, we repeat this GNC-attachment once: an existing node is selected uniformly at random; if this selects the same node as previously, this second operation aborts (\textbf{Assumption \#2 -- remix}).
    \item with probability $1-\probgnc$, $\randomnode$ is connected to $\currentnode$ (\textbf{Assumption \#3 -- refactoring}).
  \end{itemize}
\end{itemize}

The resulting graph is directed, weakly connected, and made of $N$ nodes.

\textbf{Relation to Assumption \#1 and \#2}. The first operation of the model is a node creation followed by an attachment to existing nodes using a GNC-Attach. 
This represents the creation of a new class implementing a new feature. This new feature depends upon existing classes.
The point of being attached to all dependent classes of a class means that those classes are already used, depending upon each others.
If class X depends on classes A, B and C, it means that A, B and C interact together in a way that is defined by X.
When a new node $\currentnode$ is connected to X by way of a GNC-Attach, it is also connected to A, B and C.
In other words, \emph{the new class $\currentnode$ creates a novel interaction between A, B, and C.}
Executing GNC-Attach twice\footnote{In the model, there is never more than two groups of already interacting classes being linked from a new node (a new feature). We did experiments allowing more than 2 successive GNC-Attach in the GD-GNC main loop: this has never significantly increased the fit to real data.} reflects the fact that the new class mixes two existing groups of classes.
Those groups were already interacting together separately (as witnessed by the fact that another class depends on the classes of each group). The new call brings the two groups together to provide one class with new and useful functions.

\textbf{Relation to Assumption \#3}.
The second core operation of \algoshortname (the top-level else condition) is a reverse attachment from an existing node to the added node.
It represents a refactoring operation, where a piece of code is extracted from an existing class, in order to ease re-use and to simplify the code.
Once the refactoring is performed, the newly created class is ready for being re-used. This can happen in subsequent iterations of the algorithm with the GNC-Attach.

\begin{algorithm*}[th!]
  \KwIn{$\nbiter$ the number of iterations to execute, $\probgnc$ the probability to perform a GNC-Attach, and $\probdouble$ the probability to do a second GNC-Attach conditioned on $p$.}
  \KwOut{a digraph $\graphid$ is composed of two sets: nodes ($\nodesetid$) and directed edges ($\edgesetid$)}
  \Begin{
      $\nodesetid \gets \emptyset$\\
      $\edgesetid \gets \emptyset$\\
      \For{$i \in \{ 1, ..., \nbiter\}$}{
	Create a node $\currentnode$ and add it to $\nodesetid$\\
	\eIf{$rand() \leq \probgnc$}{
	  \FGNC{$\graphid$,$\currentnode$}\\
	  \If{$rand() \leq \probdouble$}{
	    \FGNC{$\graphid$,$\currentnode$}
	  }
	}{
	  \takerandomnode \\%\hfill	\takerandomnodemath \\
	  Add an edge from $\randomnode$ to $\currentnode$\\% \hfill	($\edgesetid \gets \edgesetid \cup \{ \edgeitem{\randomnode, \currentnode}$$\}$)
	}
      }
    }
    \caption{Iterative algorithm for the ''\algoshortname'' generative model.
      \label{algo:dblegncalgo}}
\end{algorithm*}

\textbf{Analysis}.
We note that this model never modifies existing edges: at each iteration, GD-GNC adds a single node and a set of edges. We emphasize the fact that no preferential attachment is explicitly coded in the algorithm. However, an implicit preferential attachment is still present. GNC-Attach connects a new node to an existing one, but also to all successors of this existing node. As a consequence, if a node has a high in-degree, it has a higher probability of receiving a new edge.

\textbf{Parameters}.
Two parameters are required by GD-GNC and influence the growth of the graph. $\probgnc$ determines whether the new node $\currentnode$ must be added following GNC-Attach, or is connected to by an existing node. Hence, there is approximately $(1-\probgnc)~N$ nodes that have no out-going edges (sink nodes).
$\probdouble$ is the probability of a second attach conditioned on p. Increasing the number of GNC-Attach impacts the in-degree distribution, the distribution decreases more slowly when $\probdouble$ increases. Regarding the out-degree, the convexity of the distribution increases as $\probdouble$ increases.
\algoshortname is a generalization of GNC-Attach: GNC-Attach is the special case where $\probgnc = 1$ and $\probdouble = 0$. 

\subsection{Evaluation of \algoshortname}
\label{section:maxksoperator}
We now want to determine whether \algoshortname can generate graphs that are similar to real software graphs.

\subsubsection{Comparison}

We have explored several generative models described in the related work; 
for instance, the ones proposed by Kumar et al. \cite{KumarStochastic2000}, Dorogovtsev et al. \cite{DorogovtsevStructure2000}, Vazquez \cite{VazquezKnowing2000}, Grindrod \cite{GrindrodRange2002}, Bollobás et al. \cite{BollobasDirected2003} or Chakrabarti et al. \cite{ChakrabartiRmat2004}.
We emphasize the fact that we are interested in models that may be interpreted in terms of rules of software development.
However, according to our experiments, Baxter and Frean's model \cite{BaxterSoftware2008} is the only one which gives reasonable in- and out-degree distributions. 
The other ones are discarded because the resulting degree distributions are really different from the ones we observe in our dataset.
Hence, we consider Baxter and Frean's model as a baseline.
Baxter and Frean's model encodes preferential attachment based on the out-degree of nodes. Its logic is based on edge creation and edge transfer between nodes of the graph: a transfer means that either the source or the destination of an edge is modified and points to another node of the graph.

Hence, we formulate our research question as:
\textbf{\RQTwoGNCvsBaxter}

\subsubsection{Protocol}

To address this question, we first run a parameter optimization (see below) for each model (\algoshortname\ and Baxter \& Frean) on all programs of our dataset (see Table~\ref{table:maxkstable}). Then, we generate 30 synthetic graphs with each model, using the best parameters found for fitting each program.
Finally, we compute the inverse cumulative degree distribution of each graph and we compute the fitness value according to its $\maxksoperator$ value defined by Equation~\eqref{formula:ksmaxvalue}.

\textbf{Error metric.} To statistically  determine which model generates the closest graph to the true one, we compare the Kolmogorov-Smirnov statistic $K$ (as presented on section \ref{section:ksstat}) for in-degree and out-degree cumulative degree distribution of the generated graph $\mathcal{G}$ and the real graph $\mathcal{R}$. For this purpose, we define the $\maxksoperator$ function, as shown in Equation~(\ref{formula:ksmaxvalue}), which is the maximum between the two Kolmogorov-Smirnov distances: the distance $K_{\mathcal{R}_{in}, \mathcal{G}_{in}}$ between the in-degree cumulative distribution of the artificial graph $\mathcal{G}_{in}$ and the real one $\mathcal{R}_{in}$, and likewise for the out-degree distribution (resp.\@ $K_{\mathcal{R}_{in}, \mathcal{G}_{out}}$, $\mathcal{G}_{out}$, $\mathcal{R}_{out}$).

\begin{equation}
  \maxksoperator_{\mathcal{R}, \mathcal{G}} = \max(K_{\mathcal{R}_{in}, \mathcal{G}_{in}},\ K_{\mathcal{R}_{out}, \mathcal{G}_{out}})
  \label{formula:ksmaxvalue}
\end{equation}

Combining in-degree and out-degree distributions is necessary because both distributions are intimately related to each other: considering only the in-degree distribution or the out-degree distribution would be meaningless as a good in-degree distribution does not necessarily involve a good out-degree distribution, and vice-versa.
$\maxksoperator$ is a measure of error and we aim at minimizing it (the lower the better).

\textbf{Parameter optimization.} To determine the best values of $\probgnc$ and $\probdouble$ to generate graphs as close as possible to real ones, we perform a grid-search of the space of parameters, trying each value for $\probgnc$ and $\probdouble$ between 0 and 1 with a step of 0.1. For each parameter value, we generate 30 graphs.
Then, we use the $K$ statistic to assess the fitness between the true graph and the generated one. The graph with the smallest $K$ value is the one that is mostly similar to the real graph. 

\subsubsection{Results}

\begin{table}
  \caption{Average $\delta$ error of GD-GNC (G) and Baxter \& Frean models (B) expressed in $10^{-3}$. The two last columns give the p-value determined using the Mann-Whitney test; this p-value assesses whether \algoshortname\ is significantly better than Baxter \& Frean's model. The last column gives the effect size according to Cohen's formula (C).}
  \label{table:maxkstable}
	\begin{minipage}{.5\linewidth}
		\scalebox{0.9}{
			\begin{tabular}{lrrrr}
				\toprule
				\textbf{Project}&\textbf{B}&\textbf{G}&\textbf{p}&\textbf{C}\\
				\midrule
                    a4j & 1.56 & 1.67 & 0.43 & -0.12\\
                    apbsmem & 1.50 & 1.71 & 0.00 & 0.70\\
                    at-robots2-j & 1.77 & 2.33 & 0.23 & -0.39\\
                    beanbin & 2.26 & 1.53 & 0.00 & 1.92\\
                    caloriecount & 1.47 & 1.08 & 0.00 & 1.33\\
                    corina & 1.46 & 1.79 & 0.12 & -0.40\\
                    db-everywhere & 2.16 & 1.08 & 0.00 & 1.97\\
                    dom4j & 1.65 & 1.09 & 0.00 & 1.87\\
                    echodep & 1.98 & 2.17 & 0.15 & -0.37\\
                    feudalismgame & 2.13 & 1.68 & 0.12 & 0.35\\
                    fim1 & 1.74 & 2.21 & 0.01 & -0.72\\
                    follow & 1.67 & 2.93 & 0.00 & -1.42\\
                    geo-google & 1.67 & 2.80 & 0.00 & -3.64\\
                    gfarcegestionfa & 1.96 & 2.19 & 0.00 & -1.24\\
                    glengineer & 1.93 & 1.85 & 0.01 & -0.79\\
                    heal & 1.24 & 1.71 & 0.06 & -0.56\\
                    hft-bomberman & 1.33 & 1.56 & 0.88 & -0.16\\
                    httpanalyzer & 1.67 & 2.81 & 0.00 & -3.00\\
                    ifx-framework & 1.67 & 2.11 & 0.64 & -0.36\\
                    javathena & 1.52 & 1.94 & 0.00 & -1.04\\
                    jaw-br & 1.67 & 2.45 & 0.00 & -0.76\\
                    jcvi-javacommon & 1.46 & 1.81 & 0.08 & -0.43\\
                    jdbacl & 1.67 & 2.67 & 0.00 & -2.05\\
                    jhandballmoves & 1.91 & 2.09 & 0.76 & 0.10\\
                    jigen & 1.32 & 1.20 & 0.03 & 0.61\\
				\bottomrule
			\end{tabular}
		}
	\end{minipage}
	\begin{minipage}{.5\linewidth}
		\scalebox{0.9}{
			\begin{tabular}{lrrrr}
				\toprule
				\textbf{Projet}&\textbf{B}&\textbf{G}&\textbf{p}&\textbf{C}\\
				\midrule
                    jiggler & 1.89 & 1.17 & 0.00 & 2.06\\
                    jiprof & 1.20 & 1.40 & 0.01 & -0.76\\
                    jmca & 2.16 & 1.64 & 0.00 & 0.87\\
                    jnfe & 1.94 & 2.24 & 0.52 & -0.27\\
                    jsecurity & 1.84 & 1.70 & 0.02 & 0.63\\
                    jtailgui & 1.82 & 1.67 & 0.22 & -0.21\\
                    jwbf & 1.66 & 1.15 & 0.13 & 0.53\\
                    lagoon & 1.40 & 1.45 & 0.91 & 0.04\\
                    lhamacaw & 1.66 & 1.14 & 0.00 & 1.44\\
                    lilith & 2.17 & 2.30 & 0.00 & -0.86\\
                    lotus & 1.27 & 1.30 & 0.11 & 0.30\\
                    nutzenportfolio & 1.15 & 1.49 & 0.34 & -0.34\\
                    objectexplorer & 1.72 & 1.79 & 0.44 & -0.26\\
                    openhre & 1.80 & 2.03 & 0.00 & -0.84\\
                    openjms & 1.18 & 1.28 & 0.51 & 0.13\\
                    petsoar & 1.24 & 1.39 & 0.11 & -0.39\\
                    quickserver & 1.67 & 2.68 & 0.00 & -1.40\\
                    sbmlreader2 & 1.60 & 1.79 & 0.22 & -0.33\\
                    schemaspy & 1.86 & 1.97 & 0.02 & -0.63\\
                    summa & 1.96 & 1.39 & 0.00 & 2.78\\
                    twfbplayer & 1.37 & 1.49 & 0.63 & -0.16\\
                    water-simulator & 1.59 & 1.76 & 0.08 & -0.43\\
                    wheelwebtool & 2.08 & 2.03 & 0.48 & -0.10\\
                    xbus & 2.57 & 1.05 & 0.00 & 2.60\\
                    xisemele & 1.90 & 2.58 & 0.02 & -0.76\\
				\bottomrule
			\end{tabular}
		}
	\end{minipage}
\end{table}

For each software in our dataset, 
Table \ref{table:maxkstable} gives the average fit error $\maxksoperator$ of Baxter \& Frean's model (column labeled \textbf{B}) and GD-GNC (column labeled \textbf{G}). The column labeled \textbf{p} gives the p-value determined by the Mann-Whitney test; this p-value assesses whether one model is significantly better than the other one. The last column labeled \textbf{C} gives the effect size according to Cohen's formula.

We observe that \algoshortname in-degree and out-degree distributions are better than Baxter \& Frean's one in 18/\nbsubjects cases. 
In those cases, the \algoshortname\ algorithm tends to produce synthetic software graphs whose in-degree and out-degree distributions better fit those of real software dependency graphs.

We now move to a statistical assessment of the fit. We define our null hypothesis:

\NullAltHyp{the $\maxksoperator$ error values obtained for \algoshortname\ and the ones obtained from Baxter \& Frean model belong to an identical population}{the $\maxksoperator$ values obtained for \algoshortname\ and the ones obtained from Baxter \& Frean model belong to a different population}

Considering $p<0.01$, there are 10 subjects for which GD-GNC graphs are more similar to the real graphs than those generated using the Baxter \& Frean's, and in the other way, and there are 12 subjects for which Baxter \& Frean's model is significantly better. 
For 28 subjects, there is not enough evidence to say that one model is significantly better than the other.
To sum up, according to our experiments on the degree distributions, there are 10 subjects for which our \algoshortname\ model better models the software dependency structure.
This is a piece of evidence that our assumptions are not wrong (but certainly not a definitive answer, future work will strengthen or falsify this finding).

\subsection{Other properties}

As we explained earlier in this paper, though focusing the presentation on the degree distribution, we also studied many other graph properties. We report our key observations in this section. All analysis reported in this section have been performed in R, using the igraph library to manipulate graphs, and the fitdistrplus library to fit log-normal distributions.

\subsubsection{Scalar properties}

Regarding the size of the generated graph:

\begin{itemize}
\item Baxter and Frean's model generates graphs that have the correct number of edges, but they do not have the correct number of nodes. The number of nodes of these graphs ranges between half of the real graph, and 2.5 times larger. For a given real graph, the graphs generated with Baxter and Frean's model have their number of nodes that varies a lot. The coefficient of linear correlation between the number of nodes of the synthetic graphs and the real graphs is $0.82$. 
\item GD-GNC generates graphs with the correct number of nodes, and generally with a number of edges which is varying between half and 3 times the number of edges of the empirical graph. The coefficient of linear correlation between the number of edges of the GD-GNC graphs and the real graphs is $0.96$. 
\end{itemize}

Regarding the diameter and average shortest path length, GD-GNC generates graphs which diameter is within a factor of 2 with regards to the empirical graphs: the correlation between the diameter of the empirical graph and the diameter of the GD-GNC graphs is $0.59$. Baxter and Frean's generates graphs which diameter are typically very different from the empirical one: the correlation between the diameter of the empirical graph and the diameter of Baxter and Frean's graphs is $0.37$. The same sort of observations may be made about the expected relation between the diameter and the number of nodes $n$ of a scale-free graph: the diameter scales with $O(\log(n)/\log(\log(n)))$. GD-GNC graphs follow this relation quite well, which is not the case for Baxter and Frean's graphs. Likewise, the average shortest path length exhibit the same behavior.

Regarding the transitivity or clustering coefficient $C$, the correlation beween the value of $C$ for empirical graphs and its value for graphs generated by GD-GNC is $0.38$, the one between empirical graphs and Baxter and Frean's graphs is $0.51$.

Regarding the modularity $Q$, the correlation beween the value of $Q$ of empirical graphs and its value for graphs generated by GD-GNC is $0.72$, the one between empirical graphs and Baxter and Frean's graphs is $0.23$.

\subsubsection{Distribution properties}

Now we turn to more complex descriptions of graphs. Rather than a single real value, we consider sets of values: degree distributions, and spectra and eigenvectors to be more precise. Comparing 2 distributions is not as easy as comparing 2 numbers.

Let us first consider degree distributions. We have yet encountered the Kolmogrov-Smirnov statistics above. There are other tools that are interesting to pursue this comparison:

\begin{itemize}
\item the Jensen-Shannon divergence $JSD$ builds on the Kullback-Leibler divergence $D_{KL}(P||Q)$, an information theory concept that measures how much extra information is needed to code the distribution $P$ when the distribution $Q$ is already encoded.
Clearly, $D_{KL}$ is not symetric, and not transitive. We define $JSD (P||Q) = \frac{1}{2} D_{KL} (P||Q) + \frac{1}{2} D_{KL} (Q||P)$ which is obviously symetric. $\sqrt{JSD(P||Q)}$ is a metric which is called the Jensen-Shannon distance.

$JSD$ is positive, always has a finite value, and the smaller, the more similar the two distributions $P$ and $Q$ are. Using the natural logarithm to compute $JSD$, we have $0 \le{} JSD(P||Q) \le{} \ln{2}$.

\item the degree distribution of nodes follows a power law or a log-normal law. The method described in \cite{powerLawFit} is used to estimate a power law coefficient $\gamma$, and the method in \cite{fitdistr} is used to estimate the parameters $\mu$ and $\sigma$ of a log-normal law.
\end{itemize}

For each generated graph, and for in-degree and out-degree distributions, we have computed the Jensen-Shannon distance between a generated graph and the real graph. On average, the $JSD$ for GD-GNC is slightly smaller than for Baxter and Frean's: for GD-GNC: $0.10, \sigma = 0.045$ for in-degree, $0.12, \sigma = 0.075$ vs.\@ $0.12, \sigma = 0.07$ for in-degree, $0.13, \sigma = 0.105$  for Baxter and Frean's. In each case, the mean is smaller for GD-GNC, as well as the standard deviation $\sigma$. Likewise, the median is slightly smaller for GD-GNC than for Baxter and Frean's graphs; the largest values for $JSD$ is however much larger for Baxter and Frean's ($0.44$ and $0.64$ for in-degree and out-degree respectively) than for GD-GNC ($0.26$ and $0.43$ respectively).

Regarding the degree distributions, both models generate graphs with degree distributions that may be considered as fitting a power law. However, log-normal distributions may also be fitted. This is a well-known issue in real graphs \cite{newman}. Both distributions are heavy tailed.  The log-likelihood of the log-normal models fitted to GD-GNC graphs is larger than the one of graphs generated with Baxter and Frean's ($125$ on average vs.\@ $14.8$ for in-degree distributions, $8.4$ vs.\@ $3.9$ for out-degrees), and closer to the one of the log-normal models fitted to the real graphs ($131$ on average for in-degree, $34$ for out-degree). While a power law has a single parameter, a log-normal has two parameters. To compare the fits between two models having different parameters and the data, one resorts to the comparison of Akaike's Information Criterion (AIC) or the Bayesian Information Criterion (BIC) \cite{aicbic}. AIC and BIC are defined as follows:

\[
  AIC = -2 {\cal L} + 2k
\]

and 

\[
  BIC = -2 {\cal L} + k\log{n}
\]

where $n$ is the number of data used to fit a model, $k$ is the number of parameters of the model, and $\cal{L}$ is the log-likelihood of the fit. One selects the model that minimizes the chosen criterion. Either AIC, or BIC, which is to be used is a subject of debates. In our case, both criteria strongly argue for a log-normal model, whether for real graphs, for Baxter and Frean's graphs, and for GD-GNC graphs.

To summarize this study on degree distributions, the Kolmogorov-Smirnov test is not really able to conclude which among GD-GNC and Baxter and Frean's model generate graphs that are more similar to the real graphs. Using the Jensen-Shannon distance, it seems that graphs generated with GD-GNC have their degree distribution which are more similar to those generated with Baxter and Frean's model. The conclusion goes in the same direction when comparing the parameters of log-normal distributions that best fit these distributions, either on real graphs and on synthetic graphs. By the way, our experiments clearly support the idea that these distributions are much better described as log-normal than power laws.

\begin{figure}
	\centering
	\includegraphics[width=8cm]{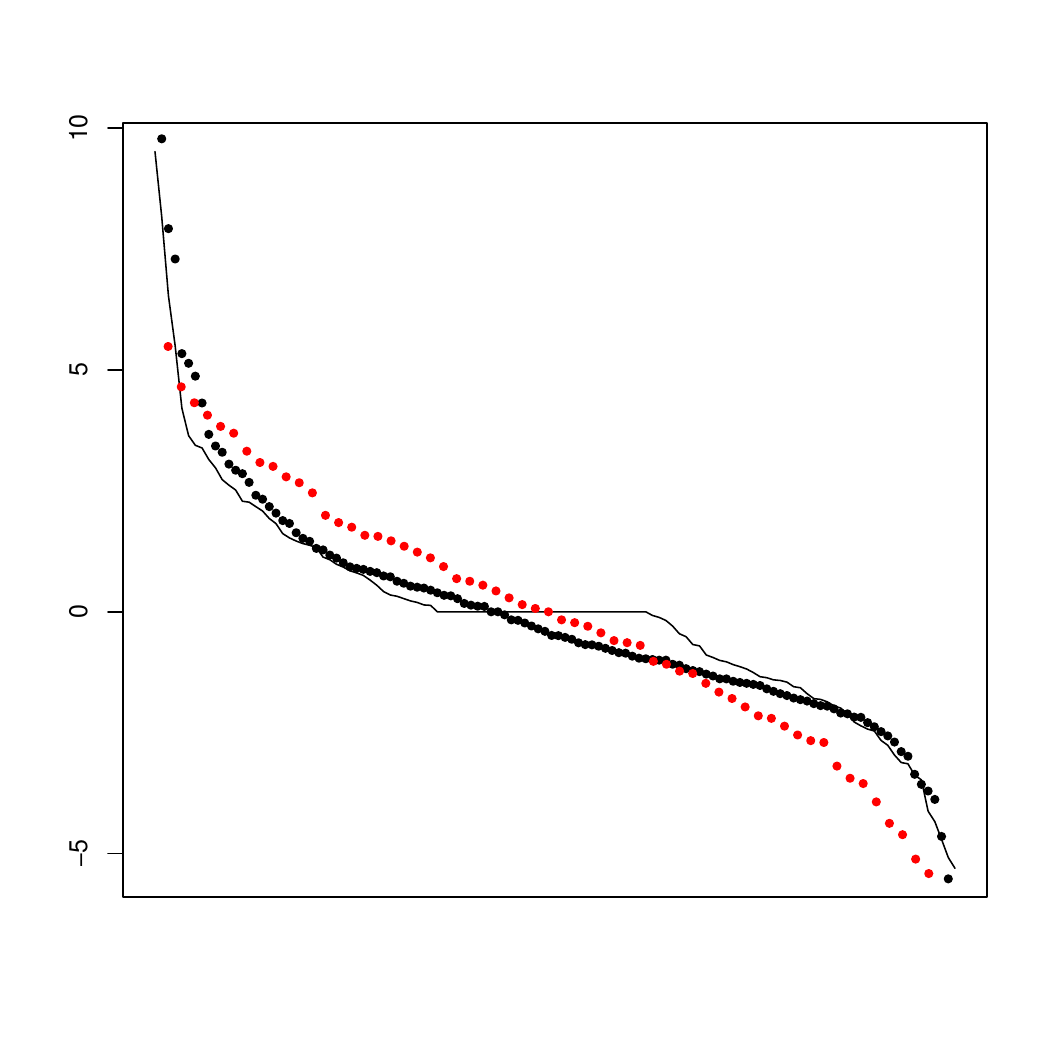}
	\caption{Example of spectra (software water-simulator was chosen because of the size of the graph): the thin line is the spectrum of the adjacency matrix of the undirected graph for this software (successive values are connected by a segment: we plot the spectrum in this way to improve legibility); the black dots are the eigenvalues of the graph generated with GD-GNC which is the closest (in terms of $\Delta$) to the real graph; the red dots are the eigenvalues of the graph generated with Baxter and Frean's which is the closest to the real graph.}
	\label{figure:spectre}
\end{figure}

Regarding spectra, we consider the spectrum of the adjacency matrix of the graphs. We consider the directed graph, and also the undirected graph. While considering the graph as undirected is meaningless from the application point of view, it helps because the spectrum of a directed graph is complex, while the spectrum of an undirected graph is real: it is easier to compare real numbers than complex numbers, and easier to plot them. For this reason, in the literature, it is very common to consider directed graphs as undirected when it comes to their spectrum. We compare the spectrum of the empirical graph and the spectrum of the graphs generated either by Baxter and Frean's model, or by GD-GNC. We measure the difference in terms of the root mean square distance: assume eigenvalues of the empirical graph are denoted $\lambda_i$ and those of the synthetic graph are denoted $l_i$, the distance between the two spectra is $\Delta = \sqrt{\frac{\sum_i(\lambda_i-l_i)^2}{n}}$. Overall, GD-GNC generates graphs that have a smaller distance than Baxter and Frean's, either considering directed or undireted graphs: the average is smaller for GD-GNC ($1.6$ vs.\@ $2.2$ for the average for the directed graphs). The fact that Baxter and Frean's graphs do not have the same number of nodes makes the comparison awkward: the problem of comparing 2 spectra of different sizes does not have a standard and recognized answer. Figure \ref{figure:spectre} illustrates the comparison of spectra: this plot shows the spectrum of the real graph as a line (in fact, this is a set of points, but we draw it as a line to ease the visual comparison with the 2 other spectra), the spectrum of the graph generated by GD-GNC with black dots, and the spectrum of the graph generated by Baxter and Frean's model with red dots. Visually, black dots are closer to the line made of red dots.

\begin{figure}
	\centering
	\includegraphics[width=6cm]{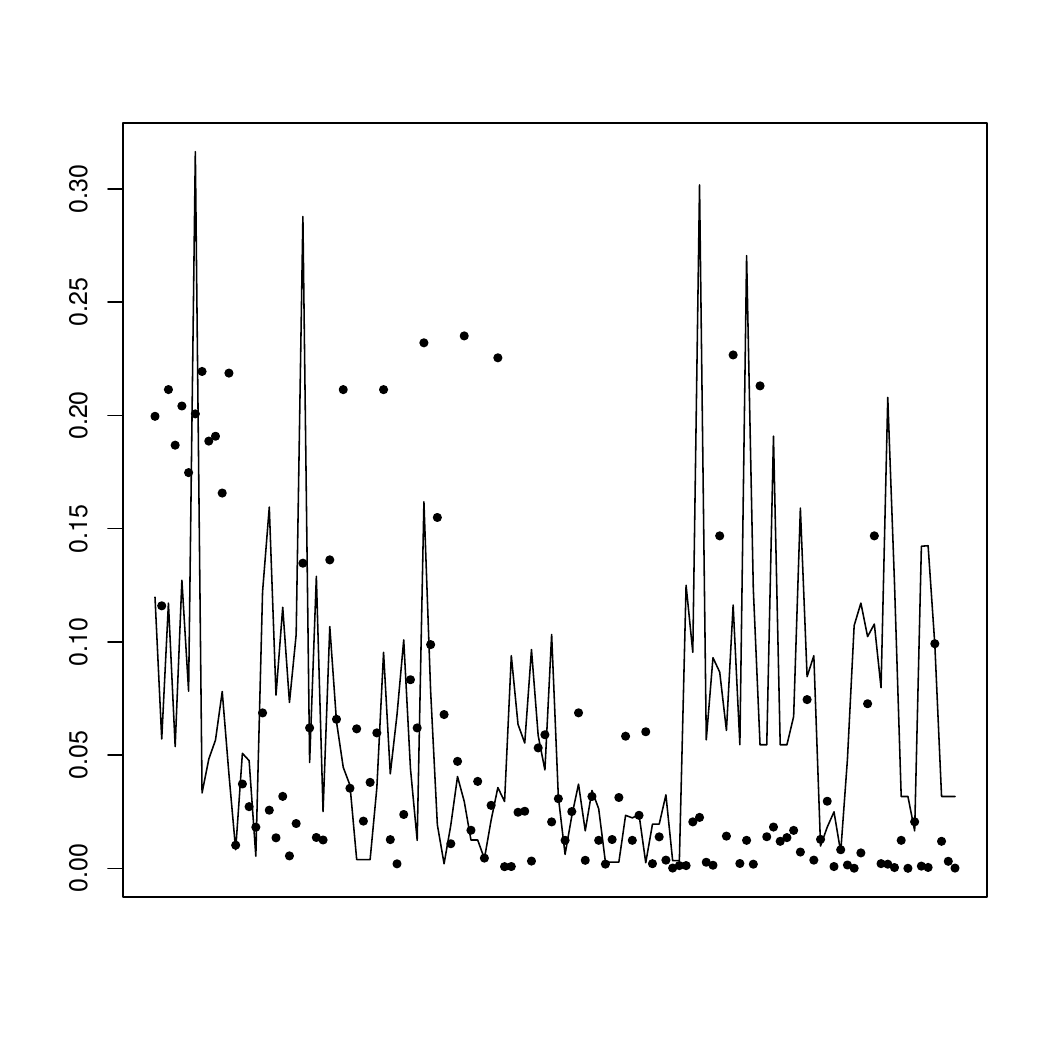}	\includegraphics[width=6cm]{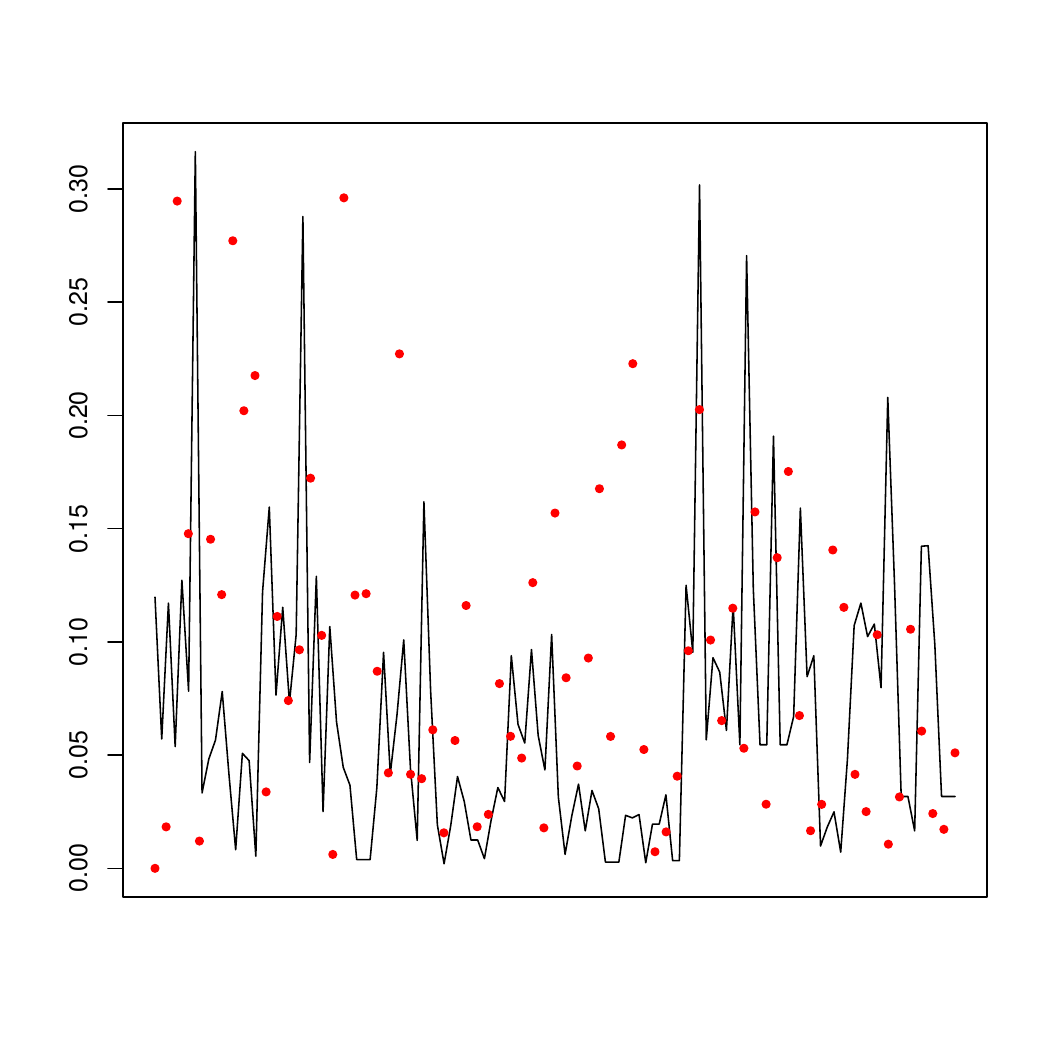}
	\caption{Example of network value. In the two plots, the line represents the network value of the real graph (successive values are connected by a segment: we plot the spectrum in this way to improve legibility). On the left, the dots is the network value of the graph generated with GD-GNC. On the right, the red dots is the network value of the graph generated with Baxter and Frean's. In each case, we consider the graph which network value is closest to the one of the real graph. We use the same software as in figure \ref{figure:spectre}.}
	\label{figure:nv}
\end{figure}

Finally, we complete our tour by considering the network value, that is the eigenvector associated to the largest eigenvalue of the spectrum of each graph. Here again, we consider graphs as undirected, as this is done in the literature. Using the same distance as for spectra $\Delta$ where the coordinates of the vectors are used instead of eigenvalues, the graphs generated by GD-GNC are again on average closer to the real graphs than those of Baxter and Frean's ($0.17$ vs.\@ $0.19$ for the average, $0.10$ vs.\@ $0.13$ for the standrard deviation, for GD-GNC and Baxter and Frean's respectively). See figure \ref{figure:nv} for a visual illustration.

\subsubsection{Summary}

Considering the different metrics we have used to compare the performance of GD-GNC with regards to other models is not an easy task. However, as reported above, most often, for each metric, the properties of graphs generated with GD-GNC are more similar to those generated by Baxter and Frean's model.

We wish to be clear about the fact that we are able to generate graphs that are even more similar to real graphs than those generated with GD-GNC. However, in this case, the model is not generating the graph incrementally, and it has no power of explanation of the rules along which the graph is generated. These graphs are of no interest for our present work, though they might prove interesting for other works related to the modeling of software graphs.

\section{Discussion}
\label{section:discussion}

We now put aside technical considerations and discuss the meaning and validity of our empirical results.

\subsection{Threats to Validity}
\label{section:threats}
Let us now discuss the threats to the validity of our findings. 
First, we have optimized our model with respect to the fit to in-degree and out-degree distributions. Even if the degree distributions capture many topological properties of graphs, it is only one feature of the structure of the dependency graph. One threat to the validity of our conclusions is that some other important topological properties of software dependency graphs have minor or no impact on degree distributions. 

Second, our experiments are done on a dataset of \nbsubjects Java software systems.
Our findings may only hold for object-oriented code, Java software or even worse, to this particular dataset only. However, for us, a sign of hope is that the degree distributions on other programming languages and systems that are reported in previous work look qualitatively the same \cite{LouridasPower2008,ValverdeLogarithmic2005,MyersSoftware2003}.

Third, our evolution model is completely expressed in abstract graph terms. We have reformulated the algorithm from a software engineering perspective in Section~\ref{section:GD-GNCmodel}. It may be the case that we have correctly extracted the core operations but that, at the same time, we have mis-interpreted their meaning. We look forward to more work in this area, to discuss with the community in order to see the emergence of a consensus on the core software evolution mechanisms. 
	
\subsection{Practical Implications}

\textbf{For Researchers} Our model and experiments have shown that remix may be a important phenomenon of software evolution. 
Our model is another piece of evidence suggesting that remix-oriented software engineering is key, strengthening existing arguments \cite{barzilay2011example,Hartmann2008}.
We note that research has already made significant advances in supporting remix of groups of classes.
For instance, code-completion can work at the level of group of classes \cite{nguyen2012graph} and documentation can be generated to explain common remix strategies \cite{Bruch2010}.
Our results call for more contributions on that topic.

\textbf{For Practitioners} The generative model is primarily intended to study hypotheses about software evolution. As such, no practitioner directly uses the model to generate new graphs. Speculatively, we envision that people who write static analysis based on dependency graphs use synthetic graphs generated by our model to validate the scalability of their technique.

\section{Related Work}
\label{section:relatedworks}

\subsection{Generic Graphs}
Several authors have proposed models for generating directed graphs in various domains. Kumar et al. \cite{KumarStochastic2000} and Bollobás et al. \cite{BollobasDirected2003} have proposed models intended to generate graphs looking like the World Wide Web graph. 
Grindrod \cite{GrindrodRange2002} has proposed a model related to protein identification on bioinformatics. 
Many other models are generic, for instance by Erdos \& Renyi \cite{ErdosRandom1959}, Dorogovtsev et al. \cite{DorogovtsevStructure2000} or Vazquez \cite{VazquezKnowing2000}, R-MAT by Chakrabarti et al. \cite{ChakrabartiRmat2004}. According to our experiments, those models are not able to reproduce software dependency graphs cumulative degree distribution. A descendant of R-MAT is the Kronecker graph model \cite{kroneckerGraphs}; however, \cite{skgICDM2011} have shown that the degree distribution of Kronecker graphs do neither fit power law nor log-normal distributions; furthermore, the generation of Kronecker can not be interpreted in terms of rules of software development.

Leskovec et al. \cite{leskovec2007graph} have proposed the forest fire model. 
This model bears some resemblance with ours.
It is guided by the idea of creating self-similar, recursive structures and the model's
``community guided attachment'' is definitely related to GNC. Their model uses so-called ``ambassador nodes'' that correspond to the selected nodes of GNC, and what they call multiple-ambassador is close to the doubling of the GNC primitive.

\subsection{Software Graphs}

Baxter \etal{} \cite{BaxterSoftware2008} studied a large amount of metrics, including graph metrics; Louridas \etal{} \cite{LouridasPower2008} studied the ``pervasive'' presence of power-law distributions on software dependencies graphs at the class and features level for a large range of software written in various languages. Myers \cite{MyersSoftware2003} also studied graph metrics on software. Nevertheless, none of them showed the common structure across software degree distributions as we have presented in this paper.
Our results on the asymmetry between the in-degree and out-degree distributions confirm previous findings by Meyers \cite{MyersSoftware2003}, Challet and Lombardoni \cite{ChalletBug2004}, Valverde and Solé \cite{ValverdeLogarithmic2005} and Baxter and Frean \cite{BaxterSoftware2008}.

The tendency of software graphs to follow a growth mechanism similar to the GNC-Attach one has been reported by Valverde and Solé \cite{ValverdeLogarithmic2005}. Moreover, they used a model of duplication and rewiring to generate software graphs with similar motifs \cite{ValverdeNetwork2005}.
Similarly, Myers \cite{MyersSoftware2003} proposed a generative model based on binary strings to materialize the software evolution rules. Unfortunately, our investigation have shown us that both models produce degree distribution that do not fit with real dependency software graphs. 
Baxter and Frean proposed a generative model of software graphs \cite{BaxterSoftware2008}, based on a preferential attachment which depends on the node degree distributions.
We have shown that our model better fits real data.

Maddison and Tarlow \cite{MaddisonStructured2014} proposed a generative model of source code. Our motivations are similar but the considered software artifacts are completely different (abstract syntax tree versus dependency graphs). 

Some authors have studied other structure characteristics of different kinds of graphs: Harman \etal \cite{HarmanDependence2009} focus on dependency clusters to demonstrate the widespread existence of clusters in software source code. They show a common traits of software using a different approach. Mitchell and Mancoridis \cite{MitchellAutomatic2006} uses clustering techniques to infer aggregate view of a software system but they want to gain understanding for specific systems in order to improve debugging and refactoring but they do not focus on generalities about software. Furthermore, none of these studies propose generative model of any kind.
Chaikalis and Chatzigeorgiou \cite{ChaikalisForecasting2015} proposed a predictive model based on a graph-based perspective at the class level. While the graphs themselves are similar to the ones studied in our work, the goal is completely different.
Chaikalis and Chatzigeorgiou's goal is to predict the growth and coupling of future versions while ours is to generate synthetic software graphs structured in a similar manner as real software graphs.

Lin and Whitehead \cite{LinWhitehead2015} have worked on a similar problem statement: they study power laws on software change size and they propose a generative model for such distributions. They work a the AST tree granularity.
However, they only focus on change sizes, where we focus on a different characteristic: the in-/out-degree distributions.

\section{Conclusion}
\label{section:conclusion}

In this paper, we have studied the fact that there is a common structure in many software dependency graphs, and devised an experimental protocol to understand the evolution principles that result in such a common structure.
Our experimental approach is to devise  a generative model of software dependency graphs and to compare the structure of artificial graphs generated with such a model against real software graphs. 

To this end, we have introduced a new generative model GD-GNC. 
This model generates graphs whose degree distribution is very close to that of real software: this closeness is assessed with statistical tests. This is a piece of evidence that the evolution rules encoded in the generative model resemble the actual ones: new features are based on the perpetual remix of existing interacting classes and refactoring mostly consists of extracting a reusable class from an existing class.

Now that we have shown that meaningful generative models exist, future work may consider properties more complex than those investigated in this paper. In particular, 
graph motifs are patterns consisting of a small amount of nodes connected to each other in a certain way.
Graph motifs \cite{MiloNetwork2002} may turn to be valuable to determine the appropriateness of generated graphs. 
We hypothesize two kinds of motifs: ones resulting from design patterns introduced at once in software, and others that are evolutionary, kind of ``emergent design patterns''. Identifying them would shed a new interesting light on software evolution. 

It is also very appealing to study a series of versions of a software. 

  \bibliographystyle{IEEEtran}
  \bibliography{paper}

\end{document}